\documentstyle[12pt,a4]{article}
\begin{document}
\begin {titlepage}
\begin{flushleft}
FSUJ TPI QO-05/98
\end{flushleft}
\begin{flushright}
March 13, 1998
\end{flushright}
\vspace{20mm}
\begin{center}
{\Large {\bf QED commutation relations for inhomogeneous 
Kramers--Kronig dielectrics}\\[3ex]
\large Stefan Scheel, Ludwig Kn\"oll, and Dirk--Gunnar Welsch}\\[2ex]
Friedrich-Schiller-Universit\"at Jena, Theoretisch-Physikalisches Institut, 
\\[1ex] Max-Wien-Platz 1, D-07743 Jena, Germany
\vspace{25mm}
\end{center}
\begin{center}
\bf{Abstract}
\end{center}
Recently a quantization scheme for the phenomenological Maxwell
theory of the full electromagnetic field in an inhomogeneous
three-dimensional, dispersive, and absorbing dielectric medium has 
been developed and applied to a system consisting of two 
infinite half-spaces with a common planar interface (H.T. Dung, 
L. Kn\"{o}ll, and D.-G. Welsch, Phys. Rev. A {\bf 57}, in press). 
Here we show that the scheme, which is based on the classical 
Green-tensor integral representation of the electromagnetic
field, applies to any inhomogeneous medium. For this purpose we
prove that the fundamental equal-time commutation relations
of QED are preserved for an arbitrarily space-dependent, 
Kramers--Kronig consistent permittivity. Further, an extension
of the quantization scheme to linear media with bounded regions
of amplification is given, and the problem of anisotropic media is briefly 
addressed. 
\end{titlepage}
\renewcommand {\thepage} {\arabic{page}}
\setcounter {page} {2}


\section{Introduction}
\label{introduction}

Quantization of the phenomenological Maxwell theory of the full
electromagnetic field in an inhomogeneous three-dimensional, 
dispersive, and absorbing dielectric medium of given permittivity 
necessarily requires a concept that is consistent with 
the principle of causality and the dissipation--fluctuation
theorem, and necessarily yields the fundamental equal-time 
commutation relations of QED. Recently it has been shown
\cite{hoknowel97} that the classical Green-tensor integral 
representation of the electromagnetic field in a medium with 
space-dependent, complex permittivity can be quantized, in
agreement with the conditions mentioned,
introducing operator noise current and charge densities and 
expressing them in terms of a continuous set of bosonic 
fields. The quantization scheme generalizes
previous work on dispersive and absorbing bulk material 
\cite{gruwel96} and one-dimensional slab-like 
systems with stepwise constant, complex permittivity 
\cite{gruwel96,gruner-welsch-1995,matloubarjef95,matlou96,%
gruner-welsch-1996b,gruner-welsch-1997}.

In particular, the ordinary vacuum QED is recognized in the limit 
when the permittivity approaches unity, and the frequently used 
approximate quantization schemes for radiation in dispersionless 
and lossless inhomogeneous media  
(see, e.g., \cite{knoell-vogel-welsch-1987,glalew91,dalton96}) 
and purely dispersive media (see, e.g., \cite{watson49,dru90,mil95})
are recognized in the narrow-bandwidth limit.
Further, the concept is in full agreement with
the Huttner--Barnett approach \cite{hutbar92} to
quantization of the electromagnetic field in bulk material. 
In this scheme, which is
based on the Hopfield model \cite{hop58} of a homogeneous dielectric, 
the electromagnetic field is coupled to a harmonic-oscillator polarization 
field that interacts with a continuous set of harmonic-oscillator reservoir
fields. All couplings are assumed to be bilinear, and the Hamiltonian of the
total system is diagonalized by using a Fano-type technique \cite{fan56}.
 
The proof of the consistence with QED of the quantization 
scheme developed in \cite{hoknowel97} requires the calculation
of some frequency integral of the (classical) Green tensor
in order to verify the fundamental equal-time commutation 
relation between the electric and magnetic fields. So far the proof for
a three-dimensional inhomogeneous medium has been based on the 
explicit expression of the Green tensor for a system that consists 
of two dispersive and absorbing bulk dielectrics with a common planar
interface. Although it is the simplest inhomogeneous system, 
the involved form of the Green tensor needs performing a
rather lengthy calculation, and the question about the
validity of the theory for more complicated three-dimensional
systems may arise. In this paper we show that the 
fundamental equal-time commutation relation between the electric
and magnetic fields is satisfied
for any inhomogeneous three-dimensional, dispersive
and absorbing dielectric medium, without making use of a 
particular form of the Green tensor. This enables us to
show that the theory applies to arbitrary inhomogeneous, linear 
media including media with bounded regions of amplification. 
Finally, we
briefly address the extension of the theory to anisotropic media.

The paper is organized as follows. The quantization scheme is outlined 
in Sec.~\ref{maxgreen}. In Sec.~\ref{properties} from the partial 
differential equation for the Green tensor an integral equation 
is derived, and general properties of the Green tensor are studied.
Section \ref{commutator} presents the proof of the fundamental
commutation relation between the electric and magnetic fields,
and in Sec.~\ref{extensions} it is shown that the scheme also
applies to media with both absorption
and (in bounded regions of space) amplification. 
Finally, a summary and some concluding remarks are given in 
Sec.~\ref{conclusions}.


\section{Quantization scheme}
\label{maxgreen}

Following \cite{hoknowel97}, we spectrally decompose the 
(Schr\"{o}\-din\-ger) electric and magnetic field operators as
\begin{equation}
\label{1}
\hat{{\bf E}}({\bf r}) = \int\limits_0^{\infty} d\omega \, 
\hat{\underline{{\bf E}}}({\bf r},\omega) +\mbox{H.c.} 
\end{equation}
and
\begin{equation}
\label{2}
\hat{{\bf B}}({\bf r}) = \int\limits_0^{\infty} d\omega \,
\hat{\underline{{\bf B}}}({\bf r},\omega) +\mbox{H.c.},
\end{equation}
respectively, where 
$\hat{\underline{{\bf E}}}({\bf r},\omega)$ and 
$\hat{\underline{{\bf B}}}({\bf r},\omega)$ satisfy Maxwell's equations
\begin{eqnarray} 
\label{3}
\nabla \!\cdot\! \hat{\underline{{\bf B}}}({\bf r},\omega) = 0, 
\end{eqnarray}
\begin{eqnarray} 
\label{4}
\nabla \!\cdot \!\left[ \epsilon_0 \epsilon({\bf r},\omega) 
\hat{\underline{{\bf E}}}({\bf r},\omega) \right] 
= \hat{\underline{\rho}}({\bf r},\omega), 
\end{eqnarray}
\begin{eqnarray} 
\label{5}
\nabla \times \hat{\underline{{\bf E}}}({\bf r},\omega) 
= i\omega \hat{\underline{{\bf B}}}({\bf r},\omega),
\end{eqnarray}
\begin{eqnarray} 
\label{6}
\nabla \times \hat{\underline{{\bf B}}}({\bf r},\omega) 
= -i\frac{\omega}{c^2} \epsilon({\bf r},\omega) 
\hat{\underline{{\bf E}}}({\bf r},\omega) +\mu_0 
\hat{\underline{{\bf j}}}({\bf r},\omega). 
\end{eqnarray}
Here, the complex-valued permittivity 
\begin{equation}
\label{7}
\epsilon({\bf r},\omega) = 
\epsilon_{\rm R}({\bf r},\omega)
+ i \epsilon_{\rm I}({\bf r},\omega) 
\end{equation}
is a function of frequency and space, with
\begin{equation}
\label{7a} 
\epsilon({\bf r},\omega) \to 1 
\quad {\rm if} \quad
\omega \to \infty.
\end{equation}
For chosen ${\bf r}$ the
real part (responsible for dispersion) and the imaginary part
(responsible for absorption) are related to each other according
to the Kramers--Kronig relations, because of causality. This also 
implies that $\epsilon({\bf r},\omega)$ is a 
holomorphic function in the upper complex frequency plane, 
\begin{equation}
\label{26}
\frac{\partial}{\partial\omega^{\ast}}\epsilon({\bf r},\omega) = 0
\quad (\omega_{\rm I} > 0).
\end{equation}
The dependence on ${\bf r}$
of $\epsilon({\bf r},\omega)$ indicates that the dielectric 
properties spatially change in general. 

In order to be consistent 
with the dissipation--fluctuation theorem, 
in Eqs.~(\ref{4}) and (\ref{6}), respectively, an operator noise  
charge density $\hat{\underline{\rho}}({\bf r},\omega)$ and an 
operator noise current density $\hat{\underline{{\bf j}}}({\bf r},\omega)$ 
have been introduced, which fulfill the equation of continuity 
\begin{equation}
\label{8}
\nabla \cdot \hat{\underline{{\bf j}}}({\bf r},\omega) = i\omega 
\hat{\underline{\rho}}({\bf r},\omega).
\end{equation}
Eventually,  
$\hat{\underline{{\bf j}}}({\bf r},\omega)$ is related
to a bosonic vector field $\hat{{\bf f}}({\bf r},\omega)$ as
\begin{equation} 
\label{9}
\hat{\underline{{\bf j}}}({\bf r},\omega) 
= \omega \, \sqrt{
\frac{\hbar\epsilon_0}{\pi} 
\,\epsilon_{\rm I}({\bf r},\omega)}
\,\hat{{\bf f}}({\bf r},\omega) ,
\end{equation}
\begin{equation}
\label{10}
\left[ \hat{f}_i({\bf r},\omega),\hat{f}^{\dagger}_j({\bf r}',\omega')\right]
= \delta_{ij} \delta({\bf r}-{\bf r}') \delta(\omega-\omega') ,
\end{equation}
\begin{equation}
\label{11}
\left[ \hat{f}_i({\bf r},\omega),\hat{f}_j({\bf r}',\omega')\right] 
= 0 =
\left[ \hat{f}^{\dagger}_i({\bf r},\omega),\hat{f}^{\dagger}_j({\bf r}',\omega')
\right] .
\end{equation}
The fields $\hat{{\bf f}}({\bf r},\omega)$ for all $\omega$
can be regarded as basic variables of an overall system 
that consists of the electromagnetic field, the polarization
field and the reservoir fields and whose Hamiltonian reads
\begin{equation}
\label{12}
\hat{H} = \int d^{3}{\bf r} \int\limits_{0}^{\infty} d\omega \,
\hat{{\bf f}}^{\dagger}({\bf r},\omega)
\!\cdot\!
\hat{{\bf f}}({\bf r},\omega).
\end{equation}   

The quantization scheme implies that the electromagnetic field
operators can be expressed in terms of $\hat{{\bf f}}({\bf r},\omega)$.
   From Maxwell's equations it is seen that 
$\hat{\underline{{\bf E}}}({\bf r},\omega)$ satisfies
the partial differential equation  
\begin{equation} 
\label{13}
\nabla \times \nabla \times \hat{\underline{{\bf E}}}({\bf r},\omega)
-\frac{\omega^2}{c^2} \epsilon({\bf r},\omega) 
\hat{\underline{{\bf E}}}({\bf r},\omega) = i \mu_0\omega
\hat{\underline{{\bf j}}}({\bf r},\omega),
\end{equation}
so that 
\begin{equation} 
\label{14}
\hat{\underline{E}}_{\,i}({\bf r},\omega)
= i \mu_0 \int d^3{\bf s} \, 
\omega G_{ik}({\bf r},{\bf s},\omega) 
\,\hat{\!\underline{j}}_k({\bf s},\omega) ,
\end{equation}
where $\,\hat{\!\underline{j}}_k({\bf s},\omega)$ is given by
Eq.~(\ref{9}), and $G_{ik}({\bf r},{\bf s},\omega)$ is the 
tensor-valued Green function of the classical problem. Here and in
the following we adopt the convention of summation over repeated
vector-component indices.
Combining Eqs.~(\ref{5}) and (\ref{14}), 
the corresponding expression for $\hat{\underline{B}}_i({\bf r},\omega)$ 
is easily derived. The integral representations of 
$\hat{E}_{i}({\bf r})$ and $\hat{B}_{i}({\bf r})$ are then
found from Eqs.~(\ref{1}) and (\ref{2}), respectively, from
which the (equal-time) commutation relations are derived to be 
\begin{equation}
\label{15}
\left[ \hat{E}_i({\bf r}),\hat{E}_k({\bf r}') \right] = 
\left[ \hat{B}_i({\bf r}),\hat{B}_k({\bf r}') \right] =0 
\end{equation}
and
\begin{equation} 
\label{16}
\left[ \hat{E}_i({\bf r}),\hat{B}_k({\bf r}') \right] 
= \frac{\hbar}{\pi\epsilon_0} \, \epsilon_{kmj} \partial_m^{r'} 
\int\limits_{-\infty}^{+\infty} d\omega\, \frac{\omega}{c^2} 
G_{ij}({\bf r},{\bf r}',\omega)
\end{equation}
($\epsilon_{kmj}$, Levi--Civita tensor; $\partial_m^{r'}$ $\!\equiv$
$\!\partial/\partial x'_m$). 
On the other hand, from QED it is well known that 
\begin{equation} 
\label{17}
\left[ \hat{E}_i({\bf r}),\hat{B}_k({\bf r}') \right] 
= - \frac{i\hbar}{\epsilon_0} \epsilon_{ikm} \partial_m^r 
\delta({\bf r}-{\bf r}'),
\end{equation}   
which reveals that the quantization scheme is in full agreement with
QED, if in Eq.~(\ref{16}) the integral over $\omega$ yields
\begin{equation}
\label{18}
\epsilon_{kmj}\partial_{m}^{r'}
\int\limits_{-\infty}^{+\infty} d\omega\, \frac{\omega}{c^2} 
G_{ij}({\bf r},{\bf r}',\omega)
= \epsilon_{kmj}\partial_{m}^{r'}
i\pi\delta_{ij}\delta({\bf r}-{\bf r}') .
\end{equation} 
It should be pointed out that this is also the condition
for obtaining the correct commutation relations for the potentials 
and canonically conjugated momenta. 

Apart from scalar electrodynamics for slab-like systems
\cite{gruwel96,gruner-welsch-1995,matloubarjef95,matlou96,%
gruner-welsch-1996b,gruner-welsch-1997},
Eq.~(\ref{18}) has been proved correct for bulk material and 
two infinite half-spaces with a common planar
interface \cite{hoknowel97,gruwel96} 
by making use of the explicit form of the Green function. 
In what follows we show that the quantization scheme yields
the correct commutation relations for arbitrary inhomogeneous 
dielectrics, i.e., for any permittivity $\epsilon({\bf r},\omega)$. 
For this purpose, let us first consider some general properties of the
Green function.


\section{Green function}  
\label{properties}

   From Eqs.~(\ref{13}) and (\ref{14}) it is easily seen that the tensor-valued 
Green function [matrix elements of the fundamental solution of Eq.~(\ref{13})]
satisfies the equation
\begin{equation}
\label{19}
\left[ \partial_i^r \partial_k^r -\delta_{ik} \left(
\triangle^r+\frac{\omega^2}{c^2} \epsilon({\bf r},\omega) \right) \right]
G_{kj}({\bf r},{\bf s},\omega) 
= \delta_{ij} \delta({\bf r}-{\bf s}).
\end{equation}
This equation and the boundary condition at infinity determine
the Green function uniquely. Similarly to Eq.~(\ref{13})
[together with Eq.~(\ref{14})] there are no nontrivial solutions
of the homogeneous problem. Let us consider absorbing bulk material. 
Since the Green function must vanish at infinity, absorption obviously 
prevents one from constructing a solution of the homogeneous equation 
which is different from zero at finite space points. 
When the dielectric material extends only over a finite region 
of space, we may assume that \mbox{$\epsilon({\bf r},\omega)$ $\!\to$
$\!1$} for \mbox{${\bf r}$ $\!\to$ $\infty$}. To preserve the analytical
properties of $\epsilon({\bf r},\omega)$, the limit 
\mbox{${\bf r}$ $\!\to$ $\infty$} must be performed first, thus 
keeping a (small) imaginary part $\epsilon_{\rm I}({\bf r},\omega)$ 
in the permittivity, which again implies that there is only the 
trivial (zero) solution of the homogeneous equation. 

It is well known that the Fourier transform of a response function 
that describes a causal relation between two physical quantities
is a holomorphic function in the upper complex frequency half-plane
(see, e.g., \cite{abrikosov,nussenzveig,altdexnussmi72}).
A typical example is the causal relation between the averages of
polarization and electric-field strength. Obviously, 
$\underline{D}_{\,ij}({\bf r},{\bf s},\omega)$ $\!=$
$\!i\mu_{0}\omega G_{ij}({\bf r},{\bf s},\omega)$ as a 
function of $\omega$ is nothing but the Fourier transform
of the tensor response function $D_{ij}({\bf r},{\bf s},\tau)$
that causally relates the electric field $E_{i}({\bf r},t)$ 
observed at space-point ${\bf r}$ and time $t$ to an 
external (point-like) current $j_{j}^{\rm ext}({\bf r},t)$
$\!=$ $J_{j}^{\rm ext}({\bf s},t)\delta({\bf r}$ $\!-$ $\!{\bf s})$
at space-point ${\bf s}$ and time \mbox{$t$ $\!-$ $\!\tau$} 
($\tau$ $\!\ge$ $\!0$) [cf. Eq.~(\ref{14})]: 
\begin{equation}
\label{25d}
E_{i}({\bf r},t)
= \int_{0}^{\infty} {d}\tau \, D_{ij}({\bf r},{\bf s},\tau) 
J_{j}^{\rm ext}({\bf s},t\!-\!\tau) .
\end{equation}
Hence, 
\begin{eqnarray} 
\label{25c}
 i\mu_{0}\omega G_{ij}({\bf r},{\bf s},\omega)
& = &\underline{D}_{\,ij}({\bf r},{\bf s},\omega)
\nonumber \\
& = & \int_{0}^{\infty} {d}\tau \, e^{i\omega\tau}D_{ij}({\bf r},{\bf s},\tau)
\end{eqnarray}
is a holomorphic
function of $\omega$ in the upper complex half-plane,
i.e.,
\begin{equation} 
\label{25a}
\frac{\partial}{\partial \omega^{\ast}} 
\, \omega
G_{kj}({\bf r},{\bf s},\omega) = 0
\quad  (\omega_{\rm I} > 0),
\end{equation}
with
\begin{equation} 
\label{25b}
\omega G_{kj}({\bf r},{\bf s},\omega) \to 0
\quad {\rm if} \quad
|\omega| \to \infty.
\end{equation}
Note that Eq.~(\ref{25a}) is in
full agreement with the differential equation (\ref{19})
[together with the boundary condition at infinity].
In this equation
the frequency $\omega$ is a
parameter, and we may assume that $G_{ij}({\bf r},{\bf s},\omega)$ 
as a function of $\omega$ is differentiable with
respect to $\omega$ in the upper complex half-plane. Applying
$\partial/\partial\omega^{\ast}$ to Eq.~(\ref{19}), we easily see that
\begin{equation}
\label{25}
\left[ \partial_i^r \partial_k^r \! - \!\delta_{ik} \! 
\left(\!
\triangle^r \! + \! \frac{\omega^2}{c^2} \epsilon({\bf r},\omega) 
\!\right)\right]
\!\frac{\partial}{\partial \omega^{\ast}}
\,\omega
G_{kj}({\bf r},{\bf s},\omega) = 0
\quad (\omega_{\rm I} > 0),  
\end{equation}
because of Eq.~(\ref{26}). 
Since there is
no nontrivial solution of the homogeneous problem,
we see that $\omega G_{kj}({\bf r},{\bf s},\omega)$ satisfies 
the Cauchy--Riemann equations (\ref{25a}).
 
  From the theory of partial differential equations it is known 
(see, e.g., \cite{garabedian}) that there exists an equivalent 
formulation of the problem in terms of an integral equation. 
As shown in App.~\ref{inteq}, $G_{ij}({\bf r},{\bf s},\omega)$
satisfies the integral equation
\begin{equation} 
\label{20}
G_{ij}({\bf r},{\bf s},\omega) = G_{ij}^{(0)}({\bf r},{\bf s},\omega) 
+ \int d^3{\bf v} \, 
K_{ik}({\bf r},{\bf v},\omega)
G_{kj}({\bf v},{\bf s},\omega),
\end{equation}
where
\begin{eqnarray}
\label{21} 
G_{ij}^{(0)}({\bf r},{\bf s},\omega) 
= \left[
\delta_{ij} - \partial_i^r \partial_j^s q^{-2}({\bf s},\omega)
\right] g(|{\bf r}\!-\!{\bf s}|,\omega)
\end{eqnarray}
and
\begin{eqnarray}
\label{22} 
\lefteqn{
K_{ik}({\bf r},{\bf v},\omega) =
\left[ \partial_k^v \ln q^2({\bf v},\omega) \right] 
\left[ \partial_i^r g(|{\bf r}\!-\!{\bf v}|,\omega) \right]
}
\nonumber \\ && \hspace{14ex} 
+\left[q^2({\bf v},\omega) - q_{0}^2(\omega) \right] 
g(|{\bf r}\!-\!{\bf v}|,\omega)\, \delta_{ik}. 
\end{eqnarray}
Here, the function
\begin{eqnarray}
\label{23}
g(|{\bf r}|,\omega) 
=
\frac{{\rm e}^{i q_{0}(\omega) |{\bf r}|}}{4\pi |{\bf r}|}=
\int \frac{{d}^{3}{\bf k}}{(2\pi)^3} \, 
\frac{e^{i{\bf k}\cdot{\bf r}}}{k^2-q_0^2(\omega)}
\end{eqnarray}
is introduced, where
\begin{equation} 
\label{24}
q^2({\bf r},\omega) = \frac{\omega^2}{c^2} \, \epsilon({\bf r},\omega)
\end{equation}
and
\begin{equation} 
\label{24a}
q_{0}^2(\omega)
= \frac{\omega^2}{c^2}\epsilon_0(\omega),
\end{equation}
$\epsilon_0(\omega)$ $\!\equiv$ 
$\!\overline{\epsilon({\bf r},\omega)}^{\,{\bf r}}$ being an 
appropriately space-averaged reference permittivity [for the
integral equation
with an ${\bf s}$-de\-pen\-dent reference permittivity 
$\epsilon_0({\bf s},\omega)$, see App.~\ref{inteq}]. 
Obviously, $G_{ij}^{(0)}({\bf r},{\bf s},\omega)$ 
is the Green function for a homogeneous medium 
with permittivity $\epsilon({\bf r},\omega)$ $\!\equiv$ 
$\!\epsilon_0(\omega)$. The second term on the right-hand
side in Eq.~(\ref{20}) essentially arises from the
inhomogeneities.
Note that according to the Fredholm alternative
the solution of the integral equation (\ref{20})
is unique, because of the non-existence of nontrivial solutions
of the homogeneous problem. 
   From Eqs.~(\ref{23}) -- (\ref{24a}) and Eq.~(\ref{7a}) 
it follows that the integral kernel $K_{ik}({\bf r},{\bf v},\omega)$,
Eq.~(\ref{22}), is a holomorphic function of $\omega$ in the upper 
complex half-plane, with
\begin{equation}
\label{24b}
K_{ik}({\bf r},{\bf v},\omega) \to 0
\quad {\rm if} \quad |\omega| \to \infty, 
\end{equation}
where $K_{ik}({\bf r},{\bf v},\omega)$ decreases as does
$\epsilon({\bf r},\omega)$ $\!-$ $\!1$.

Let us write the integral equation (\ref{20}) in the compact form
\begin{eqnarray}
\label{27}
      G = G^{(0)} + {\cal K}\,G ,
\end{eqnarray}
where 
\begin{eqnarray}
\label{28}
{\cal K}\,G & \equiv & ({\cal K}\,G)_{ij}({\bf r},{\bf s},\omega)
\nonumber \\ 
& = & \int d^3{\bf v} \, 
K_{ik}({\bf r},{\bf v},\omega) G_{kj}({\bf v},{\bf s},\omega) .
\end{eqnarray}
Assuming that $G$ can be found by iteration, we may write
\begin{equation}
\label{29}
      G = G^{(0)} + \sum_{n=1}^{\infty}{\cal K}^{n}\,G^{(0)}.
\end{equation}
   From Eq.~(\ref{21}) it is seen that $G_{ij}^{(0)}({\bf r},{\bf s},\omega)$
has a cubic singularity $|{\bf r}$ $\!-$ $\!{\bf s}|^{-3}$ for
${\bf r}$ $\!\to$ $\!{\bf s}$, and Eq.~(\ref{22}) reveals that the kernel 
$K_{ik}({\bf r},{\bf v},\omega)$ is only weakly singular 
(the singularity is weaker than the spatial dimension).
Hence, at least after the third iteration step the result is 
perfectly regular at ${\bf r}={\bf s}$. 


\section{Commutation relation} 
\label{commutator}

The results given in Sec.~\ref{properties} now enables us
to prove Eq.~(\ref{18}) for arbitrary inhomogeneous dielectrics.
For this purpose we first decompose the Green function into two parts,
\begin{equation}
\label{30}
G_{ij}({\bf r},{\bf s},\omega) 
= (G_1)_{ij}({\bf r},{\bf s},\omega) + (G_2)_{ij}({\bf r},{\bf s},\omega),
\end{equation}
where $(G_1)_{ij}({\bf r},{\bf s},\omega)$ 
and $(G_2)_{ij}({\bf r},{\bf s},\omega)$ satisfy the integral equations
\begin{equation}
\label{31}
      G_{\mu} = G_{\mu}^{(0)} + {\cal K}\,G_{\mu}
      \quad (\mu=1,2),
\end{equation}
with [cf. Eq.~(\ref{21})]
\begin{equation}
\label{32}
      (G_1)^{(0)}_{ij}({\bf r},{\bf s},\omega) = 
      \delta_{ij}\,g(|{\bf r}\!-\!{\bf s}|,\omega)      
\end{equation}
and  
\begin{equation}
\label{33}
      (G_2)^{(0)}_{ij}({\bf r},{\bf s},\omega) = 
      - \partial^s_j\, \partial^r_i\,q^{-2}({\bf s},\omega)
      g(|{\bf r}\!-\!{\bf s}|,\omega). 
\end{equation}
It is easily seen that $(G_2)_{ij}({\bf r},{\bf s},\omega)$ can be 
given by
\begin{equation}
\label{34}
      \left(G_2\right)_{ij}({\bf r},{\bf s},\omega) = 
      \partial^s_j \Gamma_i({\bf r},{\bf s},\omega), 
\end{equation}
where $\Gamma$ is the solution of the integral equation
\begin{equation}
\label{35}
      \Gamma = \Gamma^{(0)} + {\cal K}\,\Gamma,
\end{equation}
with
\begin{equation}
\label{36}
      \Gamma^{(0)}_{i}({\bf r},{\bf s},\omega) = 
      - \partial^r_i\,q^{-2}({\bf s},\omega)
      g(|{\bf r}\!-\!{\bf s}|,\omega). 
\end{equation}
Both $\omega(G_1)_{ij}({\bf r},{\bf s},\omega)$ and
$\omega(G_2)_{ij}({\bf r},{\bf s},\omega)$ 
are holomorphic functions of
$\omega$ in the upper complex half-plane,
with $\omega(G_\mu)_{ij}({\bf r},{\bf s},\omega)$ $\!\to$ $\!0$
if $|\omega|$ $\!\to$ $\!\infty$. 
Note that $\omega(G_2)_{ij}({\bf r},{\bf s},\omega)$ may be
singular at $\omega$ $\!=$ $\!0$.
Nevertheless, when substituting $G_{ij}({\bf r},{\bf s},\omega)$
from Eq.~(\ref{30}) [together with Eq.~(\ref{34})] back into 
Eq.~(\ref{14}), we can integrate by parts and use the equation of 
continuity (\ref{8}) to obtain
\begin{eqnarray}
\lefteqn{
\hat{\underline{E}}_{\,i}({\bf r},\omega)
=i\mu_0 \int d^3{\bf s} \,
\omega(G_1)_{ik}({\bf r},{\bf s},\omega) 
\,\hat{\!{\underline{j}}}_k({\bf s},\omega) 
} \nonumber \\ && \hspace{5ex} 
+ \,  \mu_0 \int d^3{\bf s} \, 
\omega^2\Gamma_i({\bf r},{\bf s},\omega) 
\hat{\underline{\rho}}({\bf s},\omega) .
\end{eqnarray}
Hence, $i\mu_0\omega (G_1)_{ik}({\bf r},{\bf s},\omega)$ and 
$\mu_0 \omega^2 \Gamma_i({\bf r},{\bf s},\omega)$ are the Fourier 
transforms of the response functions relating the 
electric-field strength to the
(noise) current density $\,\hat{\!\underline{j}}_k({\bf s},\omega)$
and the (noise) charge density $\hat{\underline{\rho}}({\bf s},\omega)$ 
separately. 
Obviously, $\omega^2 \Gamma_i({\bf r},{\bf s},\omega)$
is not singular at $\omega$ $\!=$ $\!0$.

Combining Eqs.~(\ref{30}) and (\ref{34}), we easily
see that the term on the right-hand side in Eq.~(\ref{16}) can 
be rewritten as, on recalling that 
$\epsilon_{kmj}\partial^{r'}_{m}\partial^{r'}_{j}(\ldots)$
$\!=$ $\!0$,
\begin{equation}
\label{37}
\epsilon_{kmj}\partial_{m}^{r'}
\int\limits_{-\infty}^{+\infty} d\omega\, \frac{\omega}{c^2} 
G_{ij}({\bf r},{\bf r}',\omega)
= \,
\epsilon_{kmj}\partial_{m}^{r'}
\int\limits_{-\infty}^{+\infty} d\omega\, \frac{\omega}{c^2} 
(G_1)_{ij}({\bf r},{\bf r}',\omega).
\end{equation}
Thus, only the noise-current response function 
$\sim\omega(G_1)_{ij}({\bf r},{\bf r}',\omega)$
contributes to the commutator (\ref{16}). We now substitute in 
Eq.~(\ref{37}) for $(G_1)_{ij}({\bf r},{\bf r}',\omega)$ the integral 
equation (\ref{31}) \mbox{($\mu$ $\!=$ $\!1$)} to obtain
\begin{eqnarray} 
\label{39}
\lefteqn{
\int\limits_{-\infty}^{+\infty} d\omega\, \frac{\omega}{c^2} 
(G_1)_{ij}({\bf r},{\bf r}',\omega)
= i\pi \delta_{ij}\delta({\bf r}-{\bf r}') 
} \nonumber \\ && \hspace{2ex}
+
\int\limits_{-\infty}^{+\infty} d\omega
\int d^{3}{\bf v}\, \frac{\omega}{c^2} 
\,K_{ik}({\bf r},{\bf v},\omega)(G_{1})_{kj}({\bf v},{\bf r}',\omega),
\end{eqnarray}
where the (bulk-material) relation \cite{gruwel96}
\begin{equation}
\label{40}
\int\limits_{-\infty}^{+\infty} d\omega\, \frac{\omega}{c^2} 
\,(G_{1})_{ij}^{(0)}({\bf r},{\bf r}',\omega)
= i\pi \delta_{ij}\delta({\bf r}-{\bf r}')
\end{equation}
has been used. 
Hence it remains to prove that the second term 
on the right-hand side in 
Eq.~(\ref{39}) vanishes [compare Eqs.~(\ref{37}) and (\ref{39}) 
with Eq.~(\ref{18})].

Since $K_{ik}({\bf r},{\bf v},\omega)$ and
$\omega(G_1)_{kj}({\bf v},{\bf s},\omega)$ are holomorphic functions 
of $\omega$ in the upper complex half-plane, the $\omega$ integral 
can be calculated by contour integration along 
a large half-circle (with $|\omega|$ $\!=$ $\!R$, $R$ $\!\to$ $\!\infty$).
To calculate this integral, we recall that for $|\omega|$ $\!\to$
$\!\infty$ both $K_{ik}({\bf r},{\bf v},\omega)$ and 
$\omega(G_1)_{kj}({\bf v},{\bf s},\omega)$ approach zero
at least as $\omega^{-1}$, and 
$\omega K_{ik}({\bf r},{\bf v},\omega)(G_1)_{kj}({\bf v},{\bf s},\omega)$ 
approaches zero at least as $\omega^{-2}$.
Hence, for $R$ $\!\to$ $\!\infty$ the contour integral vanishes
at least as $R^{-1}$, and the second term
on the right-hand side in Eq.~(\ref{39}) indeed equals zero, i.e.,
\begin{equation} 
\label{41}
\int\limits_{-\infty}^{+\infty} d\omega\, \frac{\omega}{c^2} 
\,(G_{1})_{ij}({\bf r},{\bf r}',\omega)
= i\pi \delta_{ij}\delta({\bf r}-{\bf r}'),
\end{equation}
which [together with Eq.~(\ref{37})] shows, that Eq.~(\ref{18})
is valid for arbitrary space-dependent permittivity. In other
words, the application of the quantization scheme to arbitrary
inhomogeneous dielectrics yields the correct QED commutation 
relation (\ref{17}). 


\section{Extensions of the quantization scheme}
\label {extensions}

The developed concept of quantization of the electromagnetic
field in a dispersive and absorbing background medium that
is described in terms of a spatially varying, complex
permittivity essentially rests on the following assumptions
and principles.
$(i)$ The permittivity $\epsilon({\bf r},\omega)$ of the
dielectric medium and the Green tensor $G_{ij}({\bf r},{\bf s},\omega)$
of the classical Maxwell equations are holomorphic functions of $\omega$
in the upper complex frequency half-plane, because of causality.
$(ii)$ There is no nontrivial solution of the
homogeneous Maxwell equations which satisfies the
boundary condition at infinity, i.e., the
electric and magnetic fields are uniquely determined
by their integral representations.
$(iii)$ To be consistent with the dissipation-fluctuation theorem,
noise current and noise charge densities must be introduced even if
there are no additional sources embedded in the
dielectric medium.
$(iv)$ Quantization then requires the integral representations
to be regarded as relations between operator-valued fields, where
the operator noise current density satisfies the commutation
relation
\begin{equation}
\label{42}
\left[\,\hat{\!{\underline j}}_{\,i}({\bf r},\omega),
\,\hat{\!{\underline j}}^{\dagger}_j({\bf r}',\omega')\right]
= \omega^2 \frac{\hbar\epsilon_0}{\pi}\,\epsilon_{\rm I}({\bf r},\omega)
\delta_{ij} \delta({\bf r}-{\bf r}') \delta(\omega-\omega') .
\end{equation}

So far we have assumed that
$\epsilon_{\rm I}({\bf r},\omega)$ $\!>$ $\!0$
($\omega$ $\!>$ $\!0$), as it is the case for absorbing media.
Obviously, the statements $(i)$ -- $(iv)$ remain valid, if
(in agreement with the Kramers--Kronig relations)
$\epsilon_{\rm I}({\bf r},\omega)$ $\!<$ $\!0$
($\omega$ $\!>$ $\!0$) in a bounded region of space, which corresponds
to the presence of an amplifying medium in that region. Obviously,
the commutation relation (\ref{42}) is obtained if (in that region) the 
operator noise current density $\hat{\underline{{\bf j}}}({\bf r},\omega)$ 
is related to the bosonic field $\hat{{\bf f}}({\bf r},\omega)$ as 
\begin{equation} 
\label{43}
\hat{\underline{{\bf j}}}({\bf r},\omega) 
= \omega \, \sqrt{
-\frac{\hbar\epsilon_0}{\pi} 
\,\epsilon_{\rm I}({\bf r},\omega)}
\,\hat{\bf f}^{\dagger}({\bf r},\omega) ,
\end{equation}
which reflects the well-known fact that amplification requires the roles 
of the noise creation and destruction operators to be exchanged (see, e.g.,
\cite{caves,glauber}). From 
inspection of Eqs.~(\ref{9}) and (\ref{43}), we see that the
two equations can be combined to express the operator noise current 
density associated with damping and amplification in 
terms of the bosonic field as 
\begin{equation}
\label{44}
\hat{\underline{\bf j}}({\bf r},\omega) 
= \omega \, \sqrt{
\frac{\hbar\epsilon_0}{\pi} 
\,|\epsilon_{\rm I}({\bf r},\omega)|} \,
\left[ \Theta\!\left(\epsilon_{\rm I} \right) \hat{\bf f}({\bf r},\omega) +
\Theta\!\left(-\epsilon_{\rm I} \right) \hat{\bf f}^{\dagger}({\bf r},\omega) 
\right],
\end{equation}
with $\Theta(x)$ being the unit step function [$\Theta(x)$ $\!=$ $\!1$
for \mbox{$x$ $\!>$ $\!0$}, and $\Theta(x)$ $\!=$ $\!0$ elsewhere].
Note that the operator noise charge density $\hat{\underline{\rho}}
({\bf r},\omega)$ is given by Eq.~(\ref{8}),
with $\hat{\underline{\bf j}}({\bf r},\omega)$ from Eq.~(\ref{44}).

The quantization scheme based on Eq.~(\ref{44}) can be regarded as 
the extension of the concept for amplifying, one-dimensional slab-like 
systems \cite{matlouartbarjef97,artlou98}
to arbitrary inhomogeneous media that contain bounded
regions in which amplification is realized. From the 
derivation given in Sec.~\ref{commutator} it is clearly seen 
that the fundamental QED commutation relation (\ref{17}) is satisfied
independently of the sign of $\epsilon_{\rm I}({\bf r},\omega)$.
For an absorbing medium the poles of 
$\omega (G_1)_{ij}({\bf r},{\bf s},\omega)$ as a function of
$\omega$ are in the lower 
complex half-plane. When the gain owing to amplification
(e.g., in a resonator-like equipment) tends to compensate for the 
losses, then the poles may approach the real axis and sharply peaked
resonances may be observed. Obviously, if there are poles on the
(real) axis, the $\omega$ integral must be performed along 
the axis $\omega$ $\!+$ $\!i\varepsilon$, $\varepsilon$ $\!\to$
$\!0$. Note that in such a case the model of linear 
amplification may fail, because of nonlinear saturation.

Another possible extension of the quantization scheme is the
inclusion in the theory of anisotropic media, for which
the permittivity is a symmetric, complex tensor function of $\omega$,
\begin{equation}
\label{45}
\epsilon_{ij}({\bf r},\omega)
= \epsilon_{ji}({\bf r},\omega),
\end{equation}
which also varies with space in general. As we will show in a forthcoming 
article, the quantization scheme also applies to the electromagnetic field in 
anisotropic Kramers--Kronig dielectrics. The calculation relies on a symmetry 
relation satisfied by the Green function according to the Lorentz reciprocity 
theorem \cite{reegiehemush87,ll8,ginzburg}, and the same integral 
relation (\ref{16}) for the fundamental commutator between the electric and 
magnetic field operators can be derived. 


\section{Conclusions} 
\label{conclusions}

We have studied quantization of the full electromagnetic field in linear,
isotropic, inhomogeneous Kramers-Kronig dielectrics, using the 
formalism of Green-tensor integral representation of the electromagnetic 
field, in which the electromagnetic field operators are related to
bosonic basic fields via the Green tensor of the classical problem.  
The formalism can be regarded as the natural extension of the
mode concept which only applies -- apart from vacuum QED -- to 
narrow-bandwidth fields. Basing on very general properties of the 
(classical) Green tensor, we have shown that the formalism yields 
exactly the QED equal-time commutation relations 
between the fundamental electromagnetic 
fields for any linear, isotropic dielectric medium. 

For this purpose we have derived an integral
equation for the Green tensor, the kernel function of which
describes the effect of spatially varying permittivity. From the
holomorphic properties of the Green tensor and the integral kernel as 
functions of frequency it then follows that the QED equal-time 
commutation relation (\ref{17}) between the electric and magnetic
fields is preserved, independently of the dependence on space
of the permittivity.   
Since the holomorphic properties are observed for absorbing media
as well as amplifying media, the quantization scheme applies
to any linear, isotropic, causal medium. The only condition is 
that amplification, which gives rise to a negative imaginary part of the 
permittivity, extends over bounded regions of space -- a condition that is 
physically always fulfilled. 
It is worth noting that the scheme can also be extended to anisotropic
media, as it will be shown in a forthcoming paper in detail.

In order to show that the quantization scheme is consistent with
QED, we have restricted attention to the equal-time commutation relations.
Clearly, the results can also be used for determining the
commutation relations of the (Heisenberg) electromagnetic field 
operators at different times. Recalling that 
$\hat{\bf f}({\bf r},\omega,t)$ $\!=$ $\hat{\bf f}({\bf r},\omega)
e^{-i\omega t}$, it can easily be derived that inclusion of
$\cos[\omega(t$ $\!-$ $\!t')]$ in the integral on the right-hand side
of Eq.~(\ref{16}) yields the commutator
$[ \hat{E}_i({\bf r},t),\hat{B}_k({\bf r}',t') ]$. Decomposing
the Green tensor as shown in Eq.~(\ref{30}) [together with
Eq.~(\ref{34})], we find that  
\begin{equation} 
\label{46}
\left[ \hat{E}_i({\bf r},t),\hat{B}_k({\bf r}',t') \right] 
= \frac{\hbar}{\pi\epsilon_0} \, \epsilon_{kmj} \partial_m^{r'} 
\!\!
\int\limits_{-\infty}^{+\infty} 
\!\!
d\omega\, \frac{\omega}{c^2} 
(G_1)_{ij}({\bf r},{\bf r}',\omega) \cos[\omega(t\!-\!t')],
\end{equation}
where $(G_1)_{ij}({\bf r},{\bf r}',\omega)$ satisfies
the integral equation (\ref{31}), with
$(G_1)_{ij}^{(0)}({\bf r},{\bf r}',\omega)$ from
Eq.~(\ref{32}). 
This result
can be regarded as the natural generalization of the 
well-known result of vacuum QED (see, e.g., \cite{cohen}).

\vspace{5mm}
\begin{flushleft}
{\Large{\bf Acknowledgements}}
\end{flushleft}
We thank T. Gruner, G. Sch\"afer, and E. Schmidt for valuable 
discussions.


\appendix

\section{Derivation of the integral equation} 
\label{inteq}

In order to derive an integral equation equivalent to 
the differential equation (\ref{19}),
we formally write in Eq.~(\ref{19})
\begin{equation} 
\label{A1}
\epsilon({\bf r},\omega)
= \epsilon({\bf r},\omega)+\epsilon_0({\bf s},\omega) -
\epsilon_0({\bf s},\omega) ,
\end{equation}
where $\epsilon_0({\bf s},\omega)$ is an appropriately chosen reference 
permittivity, which also satisfies the Kramers--Kronig relations. 
Hence we may rewrite 
Eq.~(\ref{19})
as
\begin{eqnarray} 
\label{A2} 
\left[\triangle^r +q_0^2({\bf s},\omega)\right]
G_{ij}({\bf r},{\bf s},\omega) &=& \,
\left[ q_0^2({\bf s},\omega)-q^2({\bf r},\omega) \right] 
G_{ij}({\bf r},{\bf s},\omega) 
\nonumber \\ && 
+ \, \partial_i^r \partial_k^r
G_{kj}({\bf r},{\bf s},\omega) -\delta_{ij} \delta({\bf r}-{\bf s}),
\end{eqnarray}
where the abbreviations
\begin{equation}
\label{A3}
q^2({\bf r},\omega)
= \frac{\omega^2}{c^2}\epsilon({\bf r},\omega), 
\quad
q_0^2({\bf s},\omega)=\frac{\omega^2}{c^2}\epsilon_0({\bf s},\omega)
\end{equation}
have been used.
Now we introduce the Green function 
\begin{equation}
\label{A4}
g(|{\bf r}|,{\bf s},\omega)
= \frac{{\rm e}^{i q_0({\bf s},\omega) |{\bf r}|}}{4\pi |{\bf r}|}
\end{equation}
[$q_0({\bf s},\omega)$ $\!=$ $\!(\omega/c)\sqrt{\epsilon_0({\bf s},\omega)}$],
which is easily proved to satisfy the differential equation
\begin{equation}
\label{A5}
\left[\triangle^r +q_0^2({\bf s},\omega)\right] 
g(|{\bf r}|,{\bf s},\omega) = -\delta({\bf r}).
\end{equation}
The Green function $g(|{\bf r}|,{\bf s},\omega)$ enables us to
convert Eq.~(\ref{A2}) into the integral equation
\begin{eqnarray}
\label{A6}
G_{ij}({\bf r},{\bf s},\omega)
&=& - \int d^3{\bf v} \, g(|{\bf r}\!-\!{\bf v}|,{\bf s},\omega)
\Big\{ \left[ q_0^2({\bf s},\omega) - \, q^2({\bf v},\omega) \right]
G_{ij}({\bf v},{\bf s},\omega)
\nonumber \\ && 
+ \, \partial_i^v \partial_k^v
G_{kj}({\bf v},{\bf s},\omega) 
- \delta_{ij} \delta({\bf v}-{\bf s})
\Big\} .
\end{eqnarray} 
Next we apply $\partial_i^r$ on 
Eq.~(\ref{19}) 
to obtain
\begin{equation}
\label{A7}
\partial_i^r q^{2}({\bf r},\omega) G_{ij}({\bf r},{\bf s},\omega)
= \partial_j^r\delta({\bf r}\!-\!{\bf s}),
\end{equation} 
from which we find that
\begin{eqnarray} 
\label{A8}
\lefteqn{
\partial_i^r G_{ij}({\bf r},{\bf s},\omega) = -q^{-2}({\bf r},\omega)
\partial_j^r \delta({\bf r}-{\bf s}) 
}
\nonumber\\ && \hspace{12ex}
- \left[ \partial_i^r \ln q^2({\bf r},\omega) \right] 
G_{ij}({\bf r},{\bf s},\omega) .
\end{eqnarray}
Substituting in Eq.~(\ref{A6}) for $\partial_i^v \partial_k^v
G_{kj}({\bf v},{\bf s},\omega)$ the result of Eq.~(\ref{A8}),
integrating by parts, and performing the integrals with 
$\delta$-functions, we derive
\begin{equation} 
\label{A9}
G_{ij}({\bf r},{\bf s},\omega) = G^{(0)}_{ij}({\bf r},{\bf s},\omega)
+ \int d^3{\bf v} \, K_{ik}({\bf r},{\bf v},{\bf s},\omega) 
G_{kj}({\bf v},{\bf s},\omega) . 
\end{equation}
Here,
\begin{eqnarray} 
\label{A10}
\lefteqn{
G^{(0)}_{ij}({\bf r},{\bf s},\omega) = 
\left[ \delta_{ij} \!- \! \partial_i^r \partial_j^s q^{-2}({\bf s},\omega)
\right] g(|{\bf r}\!-\!{\bf s}|,{\bf s},\omega)
}
\nonumber \\ && \hspace{8ex}
+ \, 
\left.
\partial_i^r \left[ \partial_j^s q^{-2}({\bf v},\omega) 
g(|{\bf r}\!-\!{\bf v}|,
{\bf s},\omega) \right]
\right|_{{\bf v}={\bf s}} ,
\end{eqnarray}
and the integral kernel reads
\begin{eqnarray} 
\label{A11}
\lefteqn
{
K_{ik}({\bf r},{\bf v},{\bf s},\omega) = 
\left[ \partial_k^v \ln q^2({\bf v},\omega) \right] 
\left[ \partial_i^r g(|{\bf r}\!-\!{\bf v}|,{\bf s},\omega) \right]
}
\nonumber \\ && \hspace{10ex}
+ \left[ q^2({\bf v},\omega) - q_0^2({\bf s},\omega) \right] 
g(|{\bf r}\!-\!{\bf v}|,{\bf s},\omega)\,\delta_{ik} .
\end{eqnarray}
It should be pointed out that the reference permittivity 
$\epsilon_0({\bf s},\omega)$ can be chosen freely in principle, 
since the exact solution of the integral equation (\ref{A9})
does not depend on $\epsilon_0({\bf s},\omega)$. 
In practice, however, it may be advantageous to choose 
$\epsilon_0({\bf s},\omega)$ such that $G_{ij}^{(0)}({\bf r},{\bf s},\omega)$
gives a sufficiently good zeroth-order approximation
of $G_{ij}({\bf r},{\bf s},\omega)$ for an approximate solution
of Eq.~(\ref{A9}). 

In the simplest case $\epsilon_0({\bf s},\omega)$ 
may be chosen to be independent of ${\bf s}$, e.g., by averaging 
$\epsilon({\bf r},\omega)$ over space,
\begin{equation}
\label{A12}
\epsilon_0({\bf s},\omega) \to \epsilon_0(\omega) 
= \overline{\epsilon({\bf r},\omega)}^{\,{\bf r}}.
\end{equation}
Obviously, in this case the Green funcion (\ref{A4}) and the 
integral kernel (\ref{A11}) become independent of ${\bf s}$
and Eqs.~(\ref{A9}) -- (\ref{A11}) reduce to Eqs.~(\ref{20}) -- (\ref{22})
[$g(|{\bf r}|,{\bf s},\omega)$ $\!\to$ $\!g(|{\bf r}|,\omega)$,
$K_{ik}({\bf r},{\bf v},{\bf s},\omega)$ $\!\to$
$\!K_{ik}({\bf r},{\bf v},\omega)$] 
together with Eqs.~(\ref{23}) -- (\ref{24a}). 


\end{document}